\documentclass[12pt, draftcls, onecolumn]{IEEEtran_WCM_11_00101}
\usepackage{setspace}
\usepackage{multirow}
\usepackage[dvips]{graphicx}
\usepackage[cmex10]{amsmath}
\usepackage{amsmath}
\usepackage{amssymb}
\usepackage{amsfonts}
\DeclareGraphicsExtensions{ps,eps,pst,pdf}
\usepackage{cite}

\begin{document}

\title{Relaying Technologies for Smart Grid Communications}

\vspace{-2em}

\author{Hongjian~Sun,~Bo~Tan,~Jing~Jiang,~John~S.~Thompson,\\Arumugam~Nallanathan,~and~H.~Vincent~Poor.
\thanks{Copyright (c) 2012 IEEE. Personal use of this material is permitted. However, permission to use this material for any other purposes must be obtained from the IEEE by sending a request to pubs-permissions@ieee.org.}
\thanks{This manuscript has been accepted to be published in IEEE Wireless Communications. Digital Object Identifier :  10.1109/MWC.2012.6393518. }
\thanks{H. Sun, and A. Nallanathan are with Center for Telecommunications Research, Department of Electronic Engineering, King's College London, London, WC2R 2LS, UK. (Email: mrhjsun@hotmail.com; nallanathan@ieee.org)}
\thanks{B. Tan and J. S. Thompson are with Joint Research Institute for Signal and Image Processing, Department of Electronic Engineering, University of Edinburgh, Edinburgh, EH9 3JL, UK. (Email: \{B.Tan, John.Thompson\}@ed.ac.uk) }
\thanks{J. Jiang is with the Centre for Communication Systems Research, University of Surrey, UK. (Email:jing.jiang@surrey.ac.uk)}
\thanks{H. V. Poor is with Department of Electrical Engineering, Princeton University, Princeton, NJ 08544, US. (Email: poor@princeton.edu)}}

\maketitle

\begin{abstract}
Wireless technologies can support a broad range of smart grid applications including advanced metering infrastructure (AMI) and demand response (DR). However, there are many formidable challenges when wireless technologies are applied to the smart gird, e.g., the tradeoffs between wireless coverage and capacity, the high reliability requirement for communication, and limited spectral resources. Relaying has emerged as one of the most promising candidate solutions for addressing these issues. In this article, an introduction to various relaying strategies is presented, together with a discussion of how to improve spectral efficiency and coverage in relay-based information and communications technology (ICT) infrastructure for smart grid applications. Special attention is paid to the use of unidirectional relaying, collaborative beamforming, and bidirectional relaying strategies.

\end{abstract}


\newpage

\section{Introduction}
\label{section0}

There is a widely-recognized need to upgrade existing electricity grids in order to improve power delivery, reduce operating costs and to support renewable energy sources. Due to the dependence of these goals on the data acquisition and control, such smart grids must combine existing electricity grids with advanced information and communications technology (ICT) infrastructure. A mature smart grid will consist of a number of applications, e.g., supervisory control and data acquisition (SCADA), advanced metering infrastructure (AMI), and demand response (DR). As different applications require distinct degrees of coverage, capacity, reliability, security, and latency, the implementation of ICT infrastructure for smart grid networks raises many challenging design issues \cite{white}.

Most smart grid applications, e.g., SCADA and AMI, should exhibit high reliability, large coverage and high security, while requiring different scales of latency and data rates, e.g., $0.1 \sim 1$ second latency and 100 Kbps data rate for SCADA, and $10 \sim 20$ seconds latency and 1 Mbps data rate for AMI. Considering neighborhood area networks (NANs), both wireline and wireless technologies can be used to meet these requirements. In the former case, power line communications (PLC) is a natural solution \cite{white}. Using PLC, relatively small equipment investment is needed because it uses existing power lines as the data transmission medium. Nevertheless, there are a number of challenges with PLC, e.g., low capacity.
In the latter case, either ZigBee or Wi-Fi can be employed to enable AMI and DR applications due to their good capacity and low transmit power \cite{chan, han}. However, because of their low transmit power levels, both technologies have limited coverage. A wireless mesh network that consists of various nodes (e.g., WiFi and ZigBee) organized in a mesh topology can enhance the coverage \cite{wmn, mesh1, mesh2, mesh3}. In addition, wireless mesh networks are inherently more reliable as they can take advantage of self-forming and self-healing network concepts.

Wireless mesh architectures are usually implemented at the network layer or the data link layer. Relevant research focuses on protocol design for transferring data between network entities \cite{mesh1, mesh2, mesh3}. The performance of a wireless mesh network depends on the quality, reliability and efficiency of communications between different nodes in the network. Taking advantage of spatial diversity, relaying technologies can improve the performance of wireless links between neighboring nodes to meet the communication quality requirements of a wireless mesh network. However, conventional relaying technologies, e.g., amplify-and-forward (AF) and decode-and-forward (DF), enhance the capacity at the expense of consuming resources, e.g., radio frequency (RF) spectrum. Inefficient use of these resources could lead to low spectral efficiency. Thus, it is critical to study advanced relaying technologies to improve the spectral efficiency while retaining the advantages of relaying transmission.

In the remainder of this article, we first analyze the challenges of wireless technologies for the smart grid, and then introduce conventional relaying transmission strategies. Some advanced relaying strategies are then discussed. The aim of this study is to identify relaying strategies that can achieve greater spectral efficiency, extended transmission range, and improved reliability.

\section{Challenges of ICT for Smart Grid}
\label{section2}

The smart grid ICT infrastructure should allow utilities to interact with their electrical devices as well as with the customers on a near real-time basis. However, for any wireless technology, there are several challenges that still need to be addressed before their deployment on the smart grid. These challenges are described in the following subsections.

\subsection{Coverage and Capacity Tradeoff}

Some smart grid applications, e.g., SCADA and AMI, require the access network to cover a large area. Unfortunately, since the interference level increases as the number of nodes increases, the coverage of a wireless network has an inverse relationship with the channel capacity; therefore there is a tradeoff between coverage and capacity in conventional direct transmission systems.

One potential solution is to use relaying technologies, whereby one long wireless link is broken into two or more shorter, lower power links. Due to the inherent broadcast nature of the communications from the source, it may be possible for one or more nodes receiving strong signals from the source to forward them to the destination. Therefore, relaying transmission is an important technique to widen the coverage and enhance the capacity~\cite{icc}.

\subsection{Reliability}

The reliability of a network can be defined in terms of its robustness, survivability, and sufficiency of its connectivity to support a prescribed level of performance. Most smart grid applications, e.g., AMI, require reliable communications paths from the customers back to the high-speed core network. However, radio propagation in wireless communications is affected by several factors, e.g., multipath fading which may result in a temporary failure of the communication due to a severe drop in the received signal-to-noise ratio (SNR). Furthermore, wireless networks may suffer disruptions caused by adverse weather conditions (e.g., thunderstorms) that could attenuate the transmission ability of the wireless network.

Using relaying technologies, the transmit signal can be passed through both the direct communication channel and the relay channel(s). With the aid of a  receiver combining strategy, the multipath fading effects can be averaged or even removed. Under adverse weather conditions, direct transmission link may be blocked. It is feasible to build alternate links using the rest of the nodes, therefore offering capabilities of self-forming and self-healing to the ICT infrastructure.

\subsection{Spectrum Issues}
\label{2.4}

RF spectrum is the lifeblood of wireless communication systems. However, current ICT for the smart grid has access to limited numbers of frequencies, which are primarily designated for SCADA and AMI. Using current transmission techniques, it is very challenging to support additional demands from certain new smart grid applications, such as video surveillance. The efficient use of RF spectrum is, therefore, a critical issue that needs to be addressed before the deployment of extensive wireless networking in the smart grid.

The spectral utilization efficiency is often measured by spectral efficiency, which is defined as the number of bits that can be communicated over a given bandwidth within a unit of time (in bits per second per Hz).
Due to the contributions of relay channels which can boost the signal strength at the destination, relaying technologies can achieve higher spectral efficiency than direct transmission.

\section{Conventional Relaying Strategies}
\label{section3}

Based on the above discussion, there exists a common need for applying relaying technologies in wireless networks.
We consider such a relay-based wireless system, where one or more relays working in the half-duplex mode are used to retransmit the signals to the destination (D). In this relay-based system, the data communication can be divided into two time slots. This is required due to the half-duplex constraint, which means that the relays are unable to receive and transmit data simultaneously. In the first time slot, the source (S) broadcasts its information to both the destination and one or more relays. In the second time slot, the relays forward the received data to the destination. By using relays, the destination could achieve much higher reliability in decoding the information from the source by taking advantage of spatial diversity.

\begin{figure}[!ht]
\centerline{\includegraphics[width=7in]{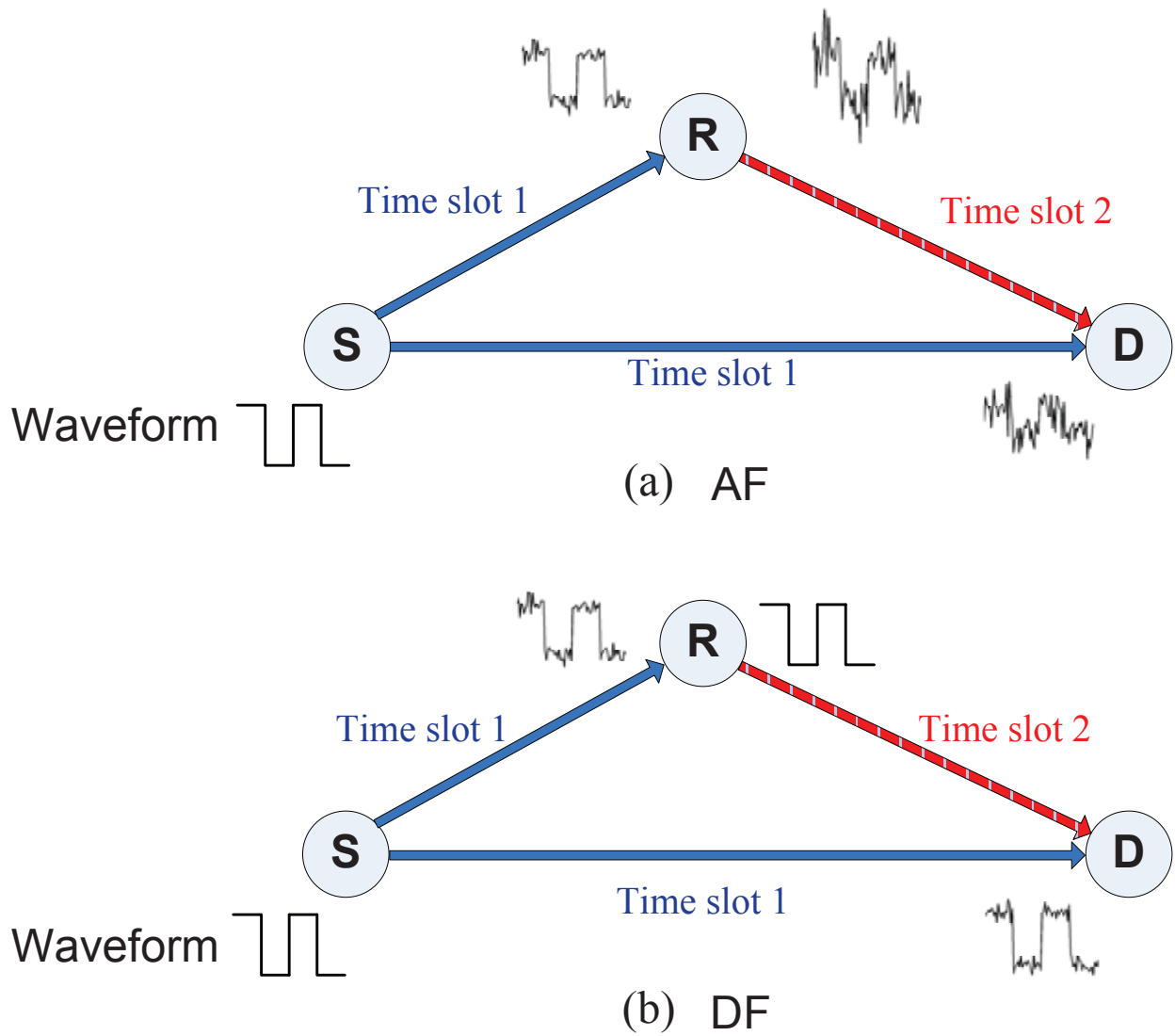}}
\caption{Demonstration of conventional relaying protocols: (a) amplify-and-forward (AF), and (b) decode-and-forward (DF).}
\label{fig2}
\end{figure}

The main challenge in the relay-based system is how to use the relays efficiently which requires study of how to use the relay(s) and also how many relays are needed. When we consider a single-relay system, the relaying protocol at the relay could significantly affect the system performance. Here, we present a brief overview of conventional relaying protocols, i.e. AF and DF. As illustrated in Fig.~\ref{fig2}(a), after the first time slot, one relay receives a noisy version of the transmitted signal from the source. The AF relaying protocol allows the relay to amplify and retransmit these noisy data to the destination \cite{af}.

Another simple relaying protocol is DF. As shown in Fig.~\ref{fig2}(b), the DF protocol allows the relay to decode the received signals from the source, and then re-encode and forward them to the destination. The performance of DF heavily depends on whether the relay can successfully decode the transmitted signals. If the relay fails to decode the signal correctly, it may be able to detect this through a cyclic redundancy check and not transmit the data. If the errors are
not detected, they will be propagated to the destination and lead to even worse performance than for direct transmission.
In either case, the relay is unable to improve detection performance at the destination. On the other hand, if the signal is correctly decoded at the relay, the destination will receive a stronger signal and thus obtain improved performance.

\section{Case Studies for Improving Spectral Efficiency of Relaying Transmission}
\label{4}

It is noteworthy that the RF spectrum in the smart grid is a very valuable resource as noted in the discussion in Section~\ref{2.4}. Unfortunately, conventional relaying technologies, i.e., AF and DF, boost the signals at the destination at the expense of consuming extra resources, e.g., the time and spectrum allocated to the relay. Inefficient use of these resources leads to low spectral efficiency. The spectral efficiency loss in a multiple-relay system (e.g., a wireless mesh network) could be even worse if either multiple time slots or frequency bands are exclusively allocated to different relays.
In order to improve the spectral efficiency while retaining the advantages of relaying transmission, it is necessary to study advanced relaying technologies. In the following subsections, we present two case studies that investigate two potential strategies, i.e. a two-relay system using beamforming concepts, and a bidirectional relaying strategy for a two-way information-exchange system.

\subsection{The Overall Setting}
\label{4.1}

\begin{figure}[!ht]
\centerline{\includegraphics[width=7in]{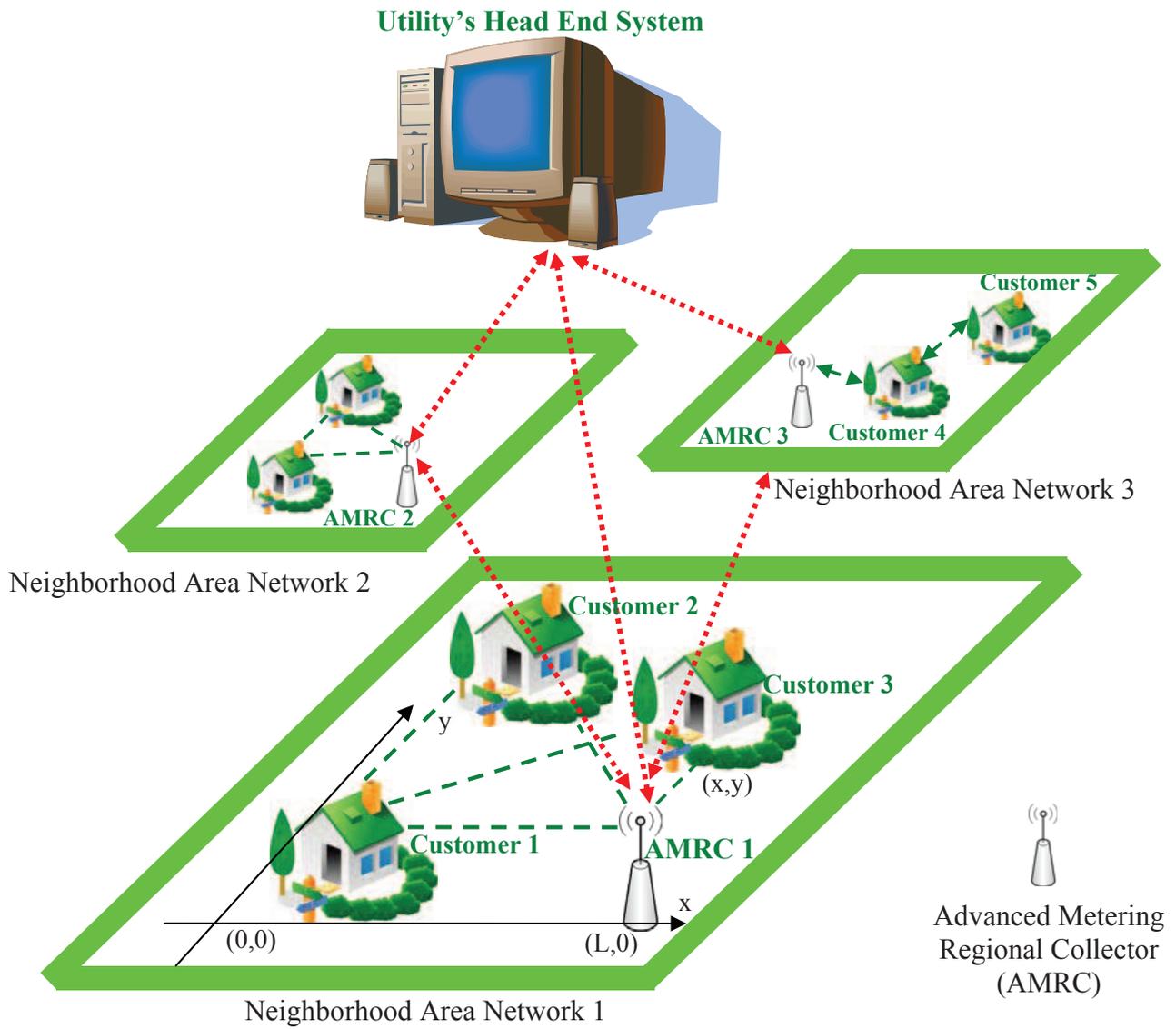}}
\caption{An example of relaying technology implementation in the smart grid: relaying technology-based advanced metering infrastructure.}
\label{fig1}
\end{figure}

Considering AMI in the smart grid, the aim is to upload energy consumption data to the utility for DR applications. An example of a wireless network for implementing AMI is shown in Fig.~\ref{fig1}. There are two scenarios in which relaying technologies can be used. In a NAN, customer~2 and customer~3 could act as relays that help customer~1 to transmit data to the advanced metering regional collector (AMRC). Further, AMRCs can help each other to transmit data to the utility's head-end system (UHES). In the following, we will discuss the implementation of relaying technologies in NAN.

Suppose that NAN is implemented using wireless technology based on the ZigBee standard, and the channel center frequency is chosen to be $2405+5(k-11)$ MHz, where the channel index $k$ is a random integer in the range $k\in[11, 26]$. The bandwidth is assumed to be $2$ MHz, and the channels between the relay and the end nodes are assumed to suffer frequency-flat Rayleigh fading. Considering the path loss, we adopt the ITU indoor propagation model \cite{model}, in which the distance power loss coefficient is set to be 28 dB/decade. The transmit power is set to be 0 dBm, and the antenna gain is 2.5 dB. Additive white Gaussian noise (AWGN) is added to the communication channels with the power level of $-110$ dBm. Without significant loss of generality, all customers are assumed to be located on the $x-y$ plane. Customer 1 is located at the origin (0,0), and the AMRC is at the coordinates $(L, 0)$. The coordinates of the relay (either customer 2 or 3) are denoted by $(x,y)$, where $x$ and $y$ are uniformly distributed values with ranges $x\in[0,L]$ and $y\in[-\frac{L}{2},\frac{L}{2}]$. To emulate the interference from other unlicensed spectrum users (e.g., WiFi or ZigBee), we assume that $1\sim 3$ users (with 3~dBm transmit power) are using the same frequency band, with random locations in the ranges $x\in[0,L]$ and $y\in[-\frac{L}{2},\frac{L}{2}]$.

\subsection{Case Study 1: Unidirectional Two-relay System with Collaborative Beamforming}
\label{4.2}

\begin{figure}[!ht]
\centerline{\includegraphics[width=7in]{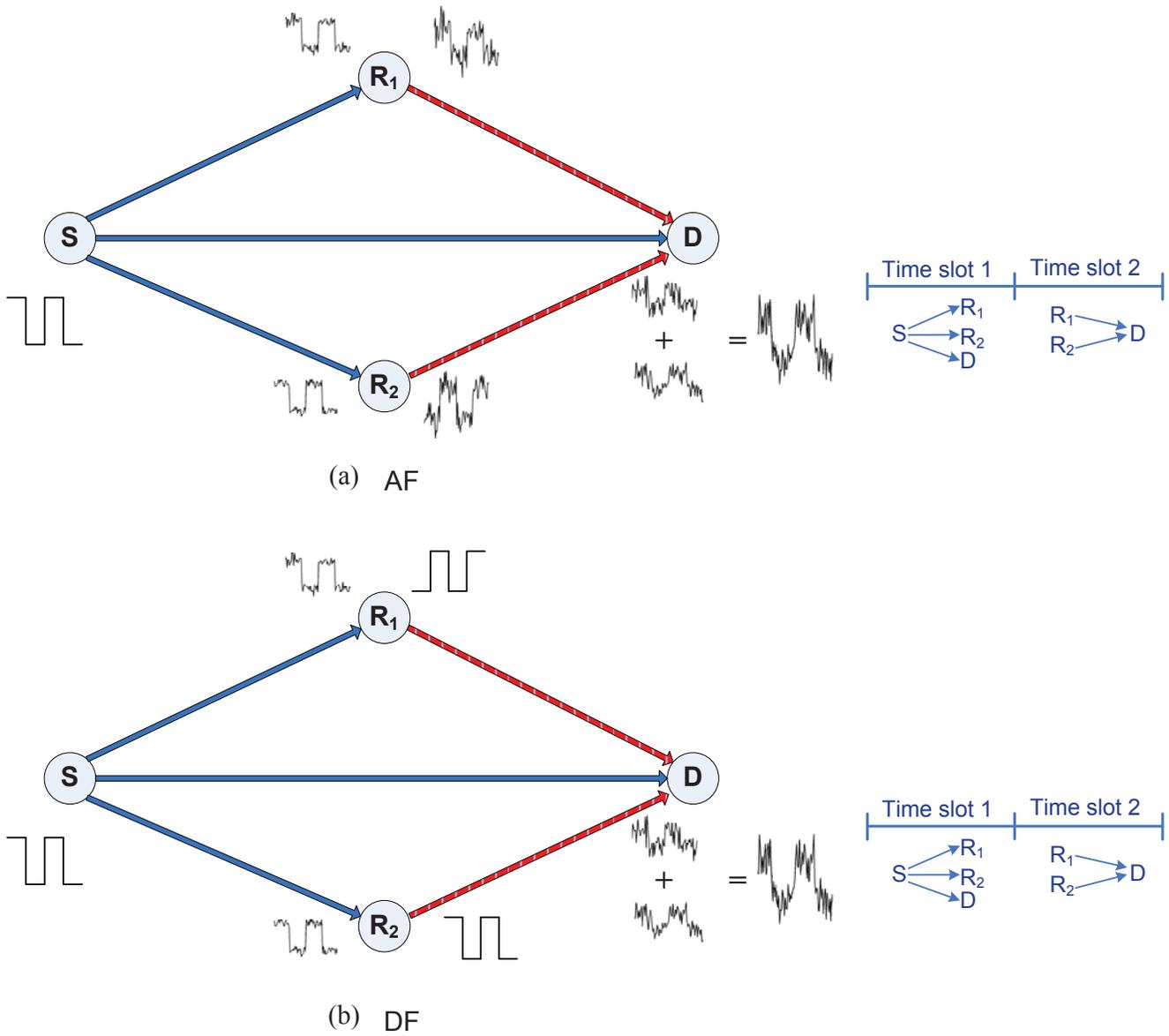}}
\caption{Relaying protocols combined with 2-relay beamforming strategy: (a) unidirectional AF relaying protocol plus beamforming, and (b) unidirectional DF relaying protocol plus beamforming. }
\label{fig3}
\end{figure}

Suppose that one smart meter of customer 1 (source) is uploading the data to the AMRC (destination) while two neighboring customers (relays) could assist the data transmission procedure. Using conventional relaying strategies, two relays forward certain versions of the received signals to the destination, as shown in Fig.~\ref{fig3}(a) for AF and Fig.~\ref{fig3}(b) for DF. No matter which relaying protocol is used, AF or DF, the two-relay system faces a challenge: Because of different channel phases at the two relays, the correlation properties of the received signals at the destination will be distorted. That means, a superposition of the signals at the destination will not necessarily strengthen the intended signals.

Collaborative beamforming \cite{poor} can be introduced to adaptively adjust the transmit signal phases and amplitudes at the two relays. For example, as shown in Fig.~\ref{fig3}, the signal phase at the relay $2$ is adjusted. Taking advantage of collaborative beamforming, the received signals at the destination can be constructively added at the destination to improve SNR.
To enable collaborative beamforming, the relays are assumed to be synchronized by the use of reference signals from a positioning system such as the global positioning system (GPS). In addition, we assume that the relays are sufficiently separated so that any mutual coupling effects among their antennas are negligible.

\begin{figure}[!ht]
\begin{center}$
\begin{array}{cc}
\includegraphics[width=92mm]{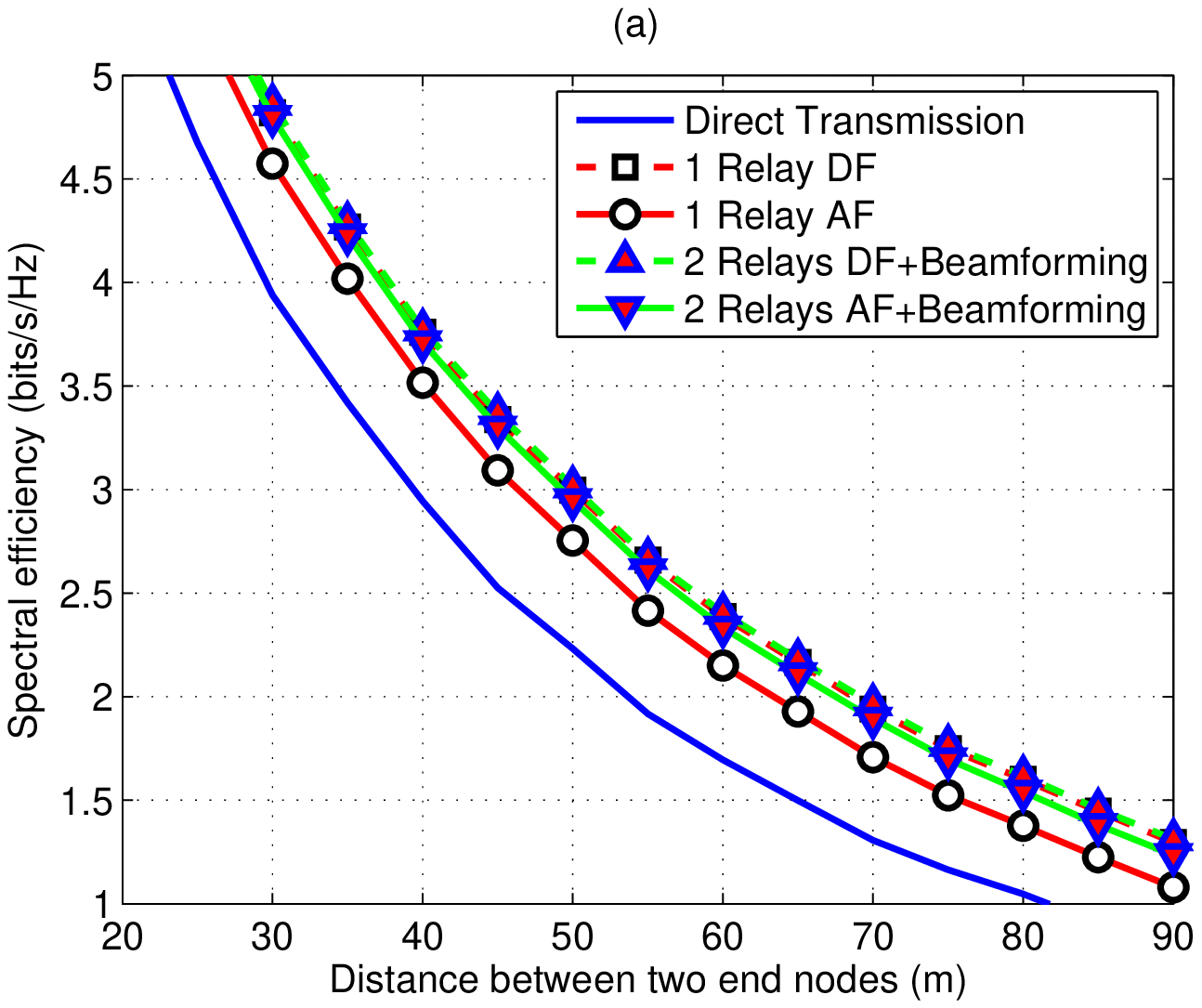}\!\!\!\!\!\! &\!\!\!\!\!\!
\includegraphics[width=92mm]{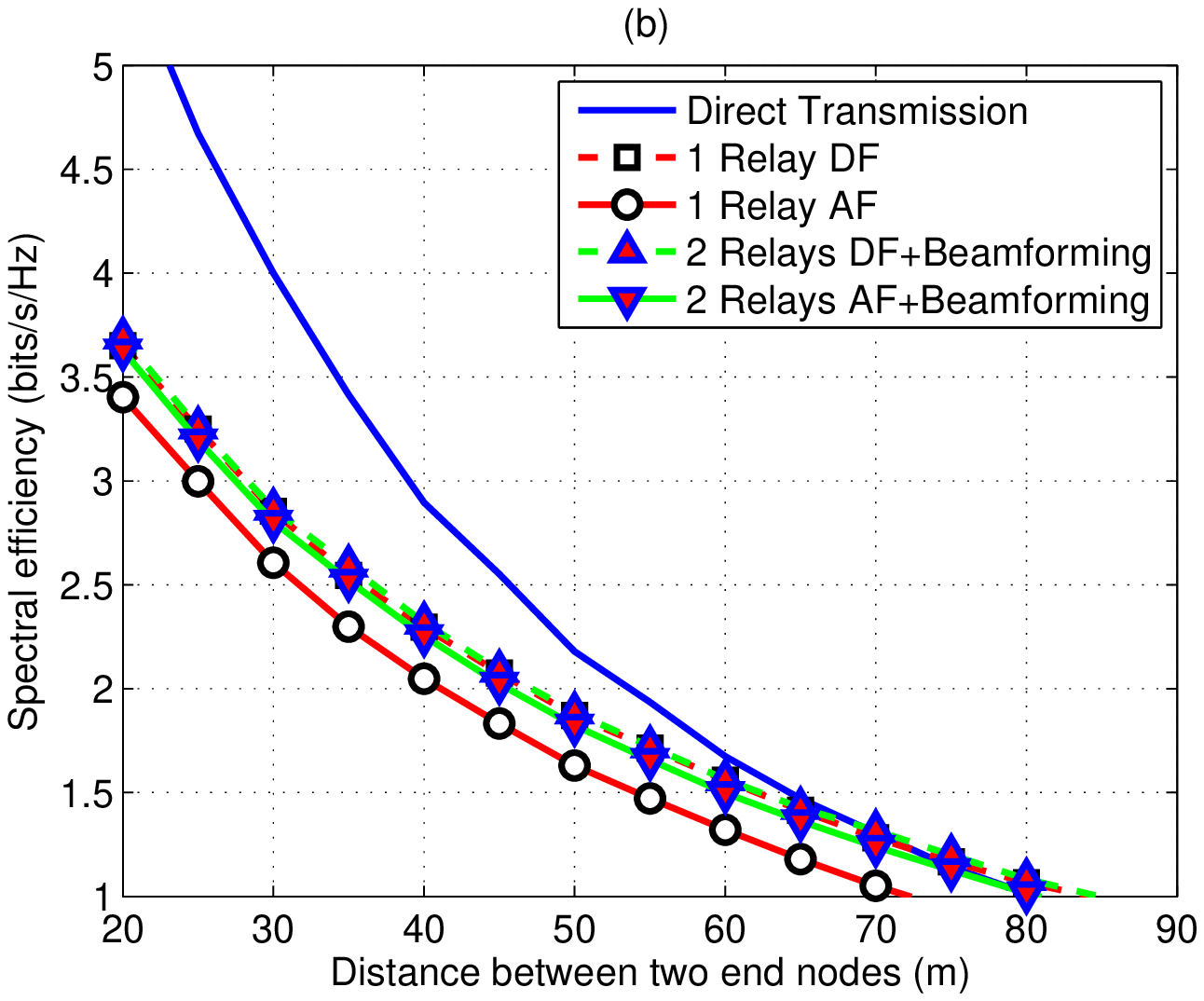} \\
\includegraphics[width=92mm]{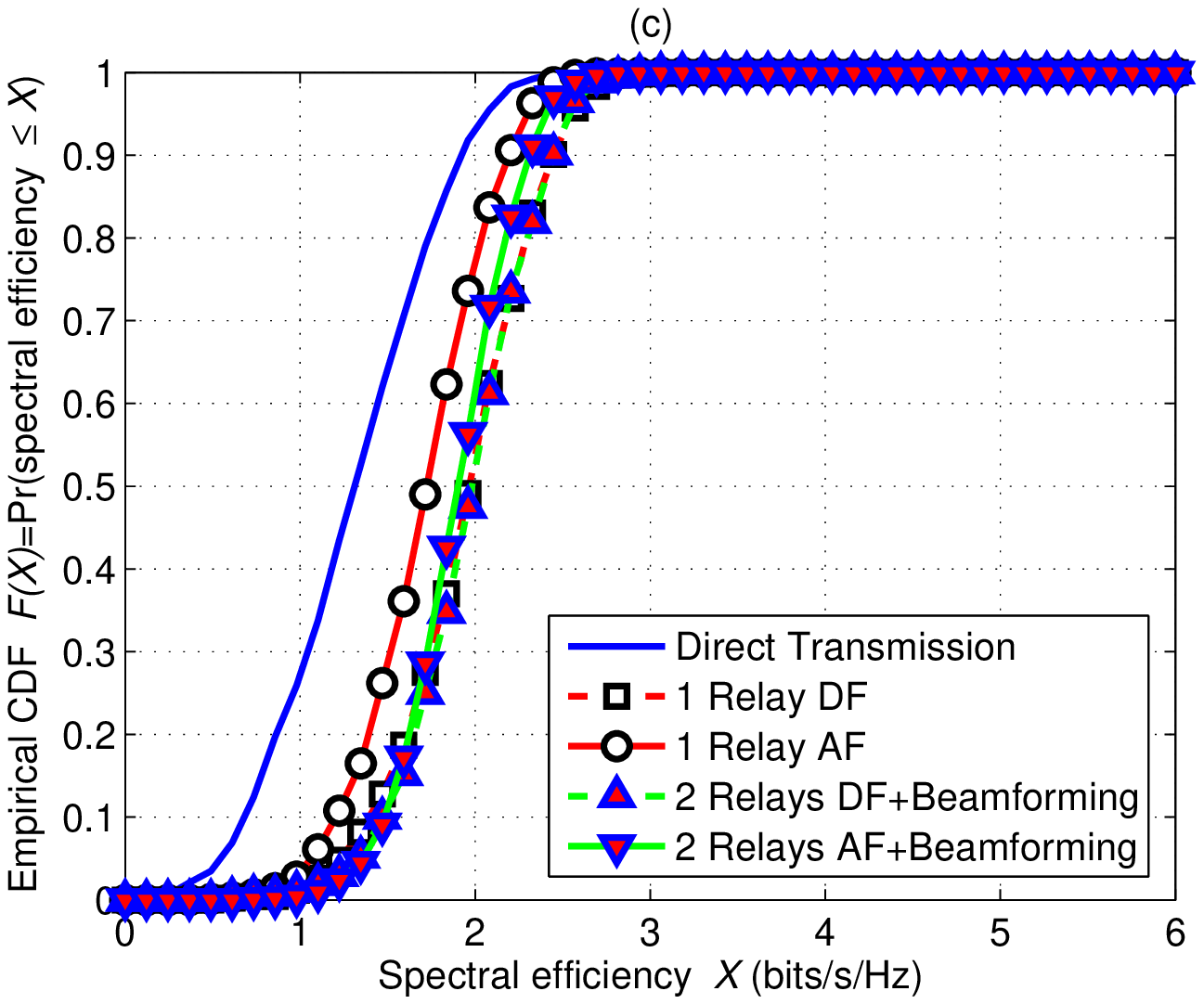}\!\!\! \!\!\!&\!\!\!\!\!\!
\includegraphics[width=92mm]{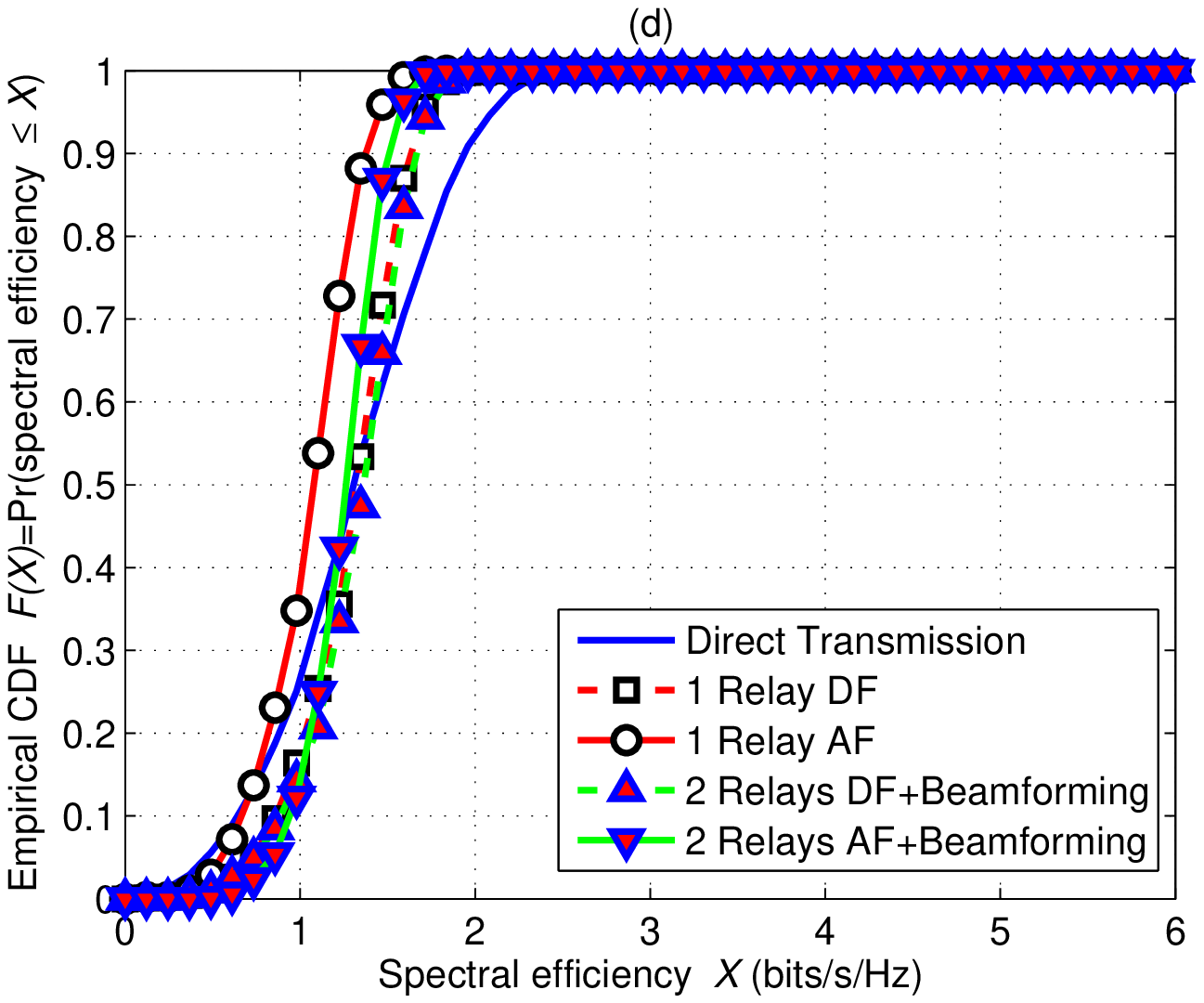}
\end{array}$
\end{center}
\caption{Comparisons of relaying technologies with direct transmission: (a) Spectral efficiency when direct transmission is available, (b) Spectral efficiency when direct transmission is blocked, (c) Empirical spectral efficiency cumulative distribution function when direct transmission is available, and (d) Empirical spectral efficiency cumulative distribution function when direct transmission is blocked.}
\label{fig4}
\end{figure}

The spectral efficiencies of different relaying strategies are compared in Fig. \ref{fig4}(a) and (b), where the $x$-axis denotes the distance between two end nodes and the $y$-axis denotes the system spectral efficiency. As shown in Fig. \ref{fig4}(a), when direct transmission is available, relaying strategies yield spectral efficiency (also coverage) gain over direct transmission for any transmission distance. Beamforming in the unidirectional two-relay AF system results in marginal improvement compared to the single-relay AF strategy, while adding an extra relay node also increases the system complexity. Meanwhile, the performance of the two-relay DF system with collaborative beamforming is similar to the single-relay DF strategy. This is because the overall information rate of the system is, in fact, limited by the channel conditions between the source and the two relays in the first time slot. One poorly conditioned channel from the source to one of the relays will eventually impair the overall spectral efficiency of the two-relay DF strategy. In Fig. \ref{fig4}(b), we can see that, if direct transmission is blocked, relaying technologies can still achieve satisfactory spectral efficiency.

The empirical cumulative distribution function (CDF) of spectral efficiency is shown in Fig.~\ref{fig4}(c) and (d), where the $x$-axis denotes the spectral efficiency $X$ and the $y$-axis denotes the empirical CDF $F(X)$. The spectral efficiency CDF $F(X)$ is defined as the percentage of systems having a spectral efficiency less than or equal to $X$. The simulation results are based on an assumption that the distance between the two end nodes is 70 metres. The CDF shows results for 10,000 different channel conditions for a given distance between the two end nodes. It should be emphasized that the steeper the curve is, the more robust the system can be. In addition, shifting the curve to the right implies that it can obtain higher spectral efficiency. In Fig. \ref{fig4}(c) and (d), we can see that all relaying strategies are more reliable than direct transmission.

\subsection{Case Study 2: Bidirectional Relaying for Information-exchange System}
\label{4.3}

DR applications in the smart grid require high data-rate two-way communications between the customers and the UHES.
Using conventional unidirectional relaying strategies in NAN, 4 time slots are needed to accomplish the information exchange process, leading to low spectral efficiency.

\begin{figure}[!ht]
\centerline{\includegraphics[width=7in]{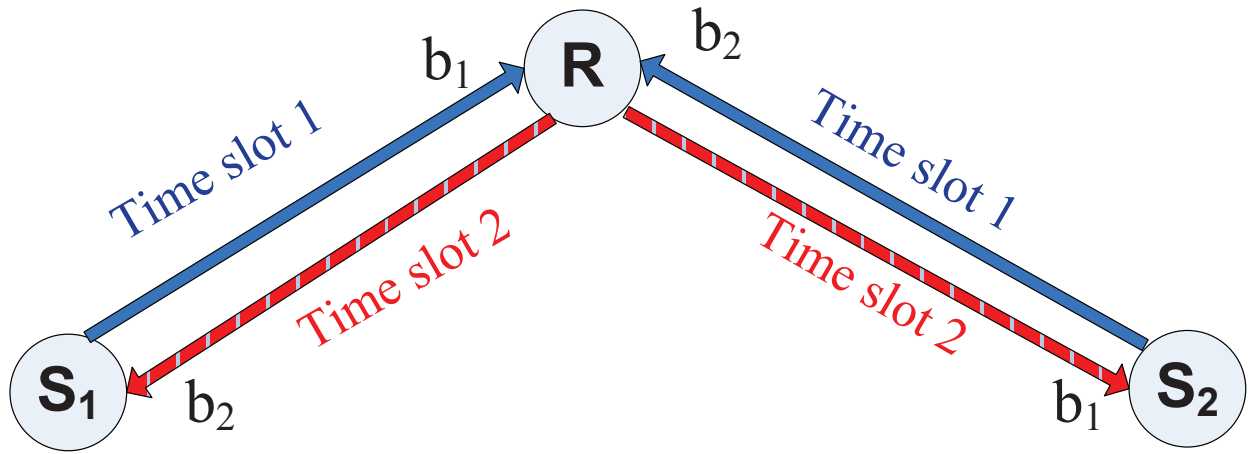}}
\caption{The bidirectional relaying strategy that only uses 2 time slots when two end nodes exchange information.}
\label{fig5}
\end{figure}

As shown in Fig.~\ref{fig5}, the bidirectional relaying strategy requires only 2 time slots to complete an information exchange process. In the first time slot, two end nodes (e.g., customer~5 and the AMRC 3 in Fig.~\ref{fig1}) send their information to the relay (e.g., customer 4) using the same frequency band, leading to a superposition of the received signals at the relay. Note that because we assume that all nodes are working in the half-duplex mode, two end nodes cannot decode the signals in the first time slot. Using the AF relaying protocol, the relay directly forwards its received signals to both end nodes. Because these two end nodes have their own copies of transmitted signals, they can subtract their own signals and obtain the information transmitted from the other node. We note that the channel state information between the relay and two end nodes can be estimated by using a channel estimation scheme, e.g., pilot symbol insertion or training bits \cite{nallan1, nallan2}. In this way, a higher spectral efficiency can be achieved, since the information at the two nodes is exchanged using fewer time slots.

If the bidirectional relaying strategy is applied together with the DF relaying protocol, the relay needs to decode the superposed signals. Optimal performance can be achieved using a maximum likelihood (ML) detection algorithm. Using ML, the detection error can be minimized but at the expense of high complexity. With lower computational cost, other approaches can obtain near ML performance. One approach is the so-called vertical Bell Labs layered space time (V-BLAST) detection algorithm \cite{blast}. Using V-BLAST, we first detect the signal from one node by treating the signal from another node as interference. We then subtract the detected signal vector from the received signal vector, and perform detection on the resulting signal vector. The advantage of the V-BLAST algorithm is that its computational complexity is low and fixed for the whole range of SNRs. Therefore, we consider the V-BLAST detection algorithm at the relay for the bidirectional DF relaying system.

\begin{figure}[!ht]
\begin{center}$
\begin{array}{cc}
\includegraphics[width=92mm]{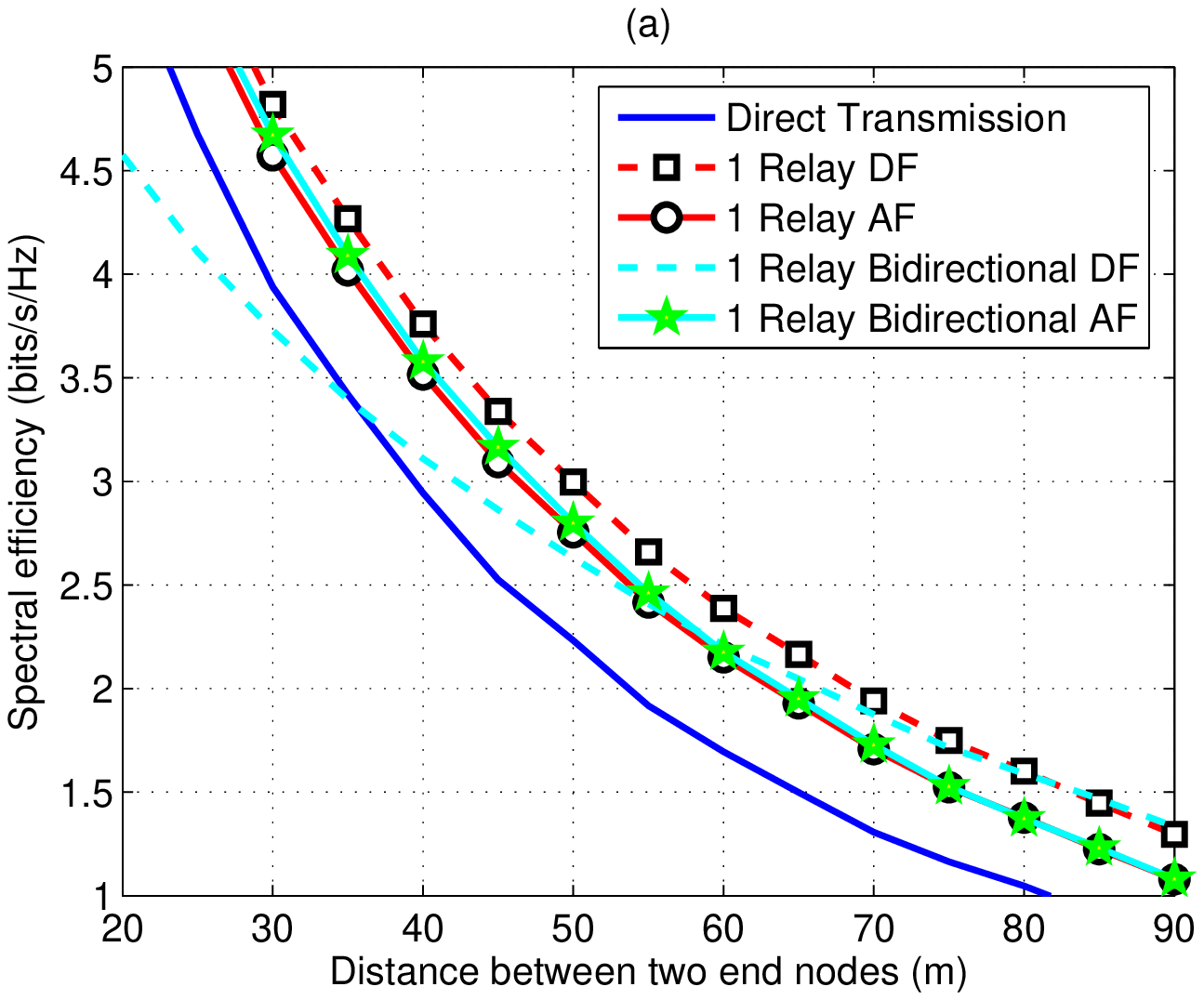}\!\!\!\!\!\! &\!\!\!\!\!\!
\includegraphics[width=92mm]{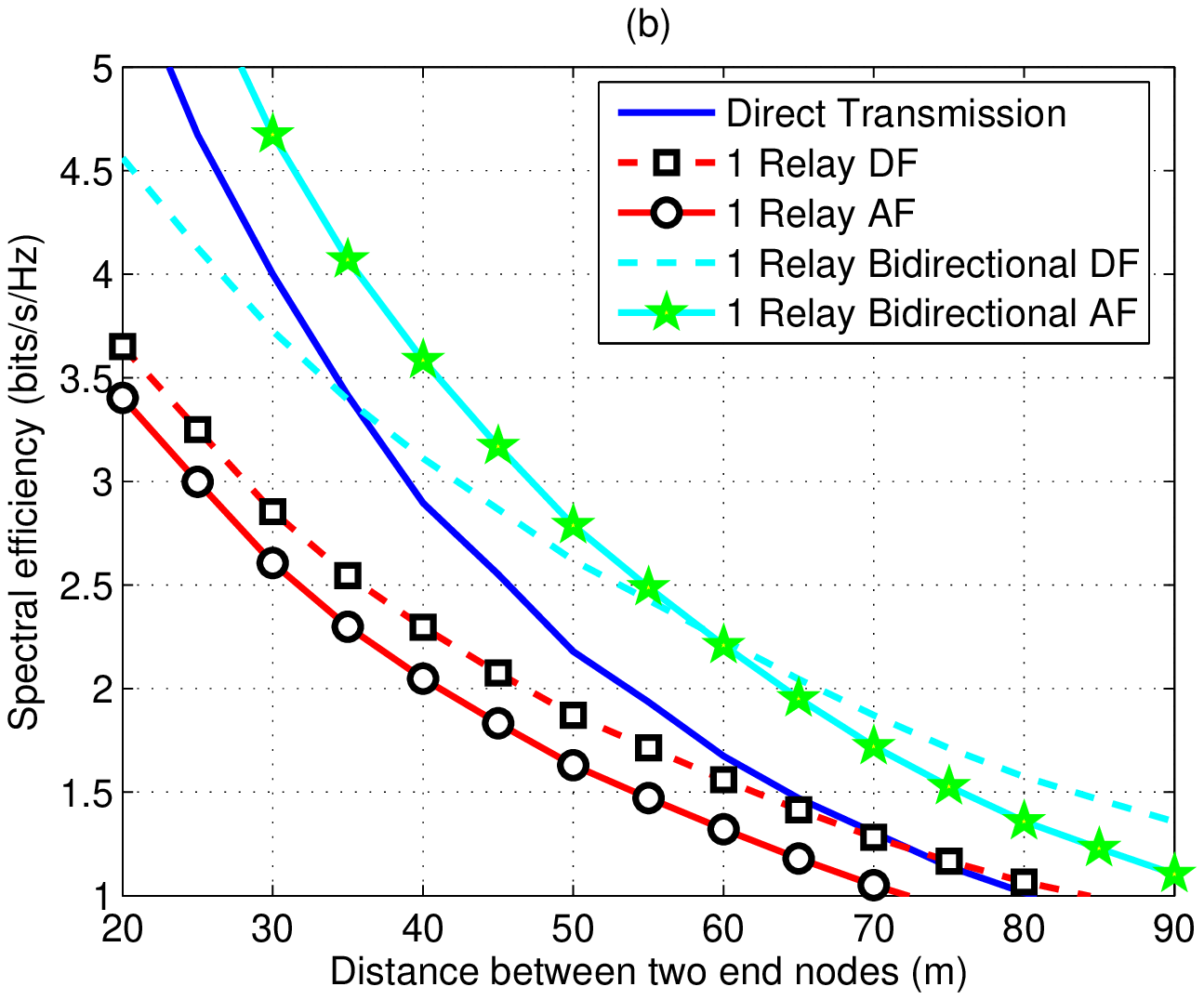} \\
\includegraphics[width=92mm]{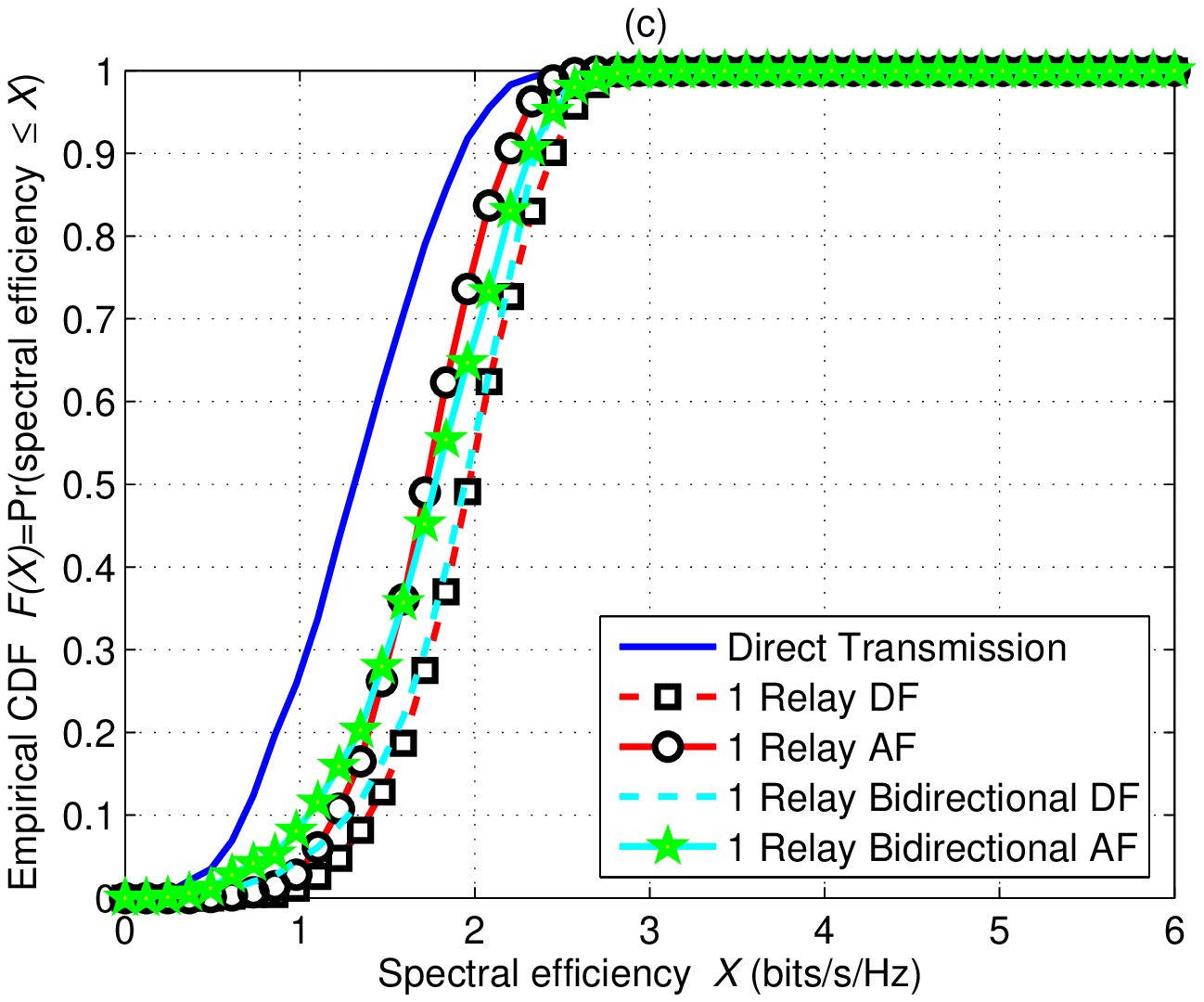}\!\!\! \!\!\!&\!\!\!\!\!\!
\includegraphics[width=92mm]{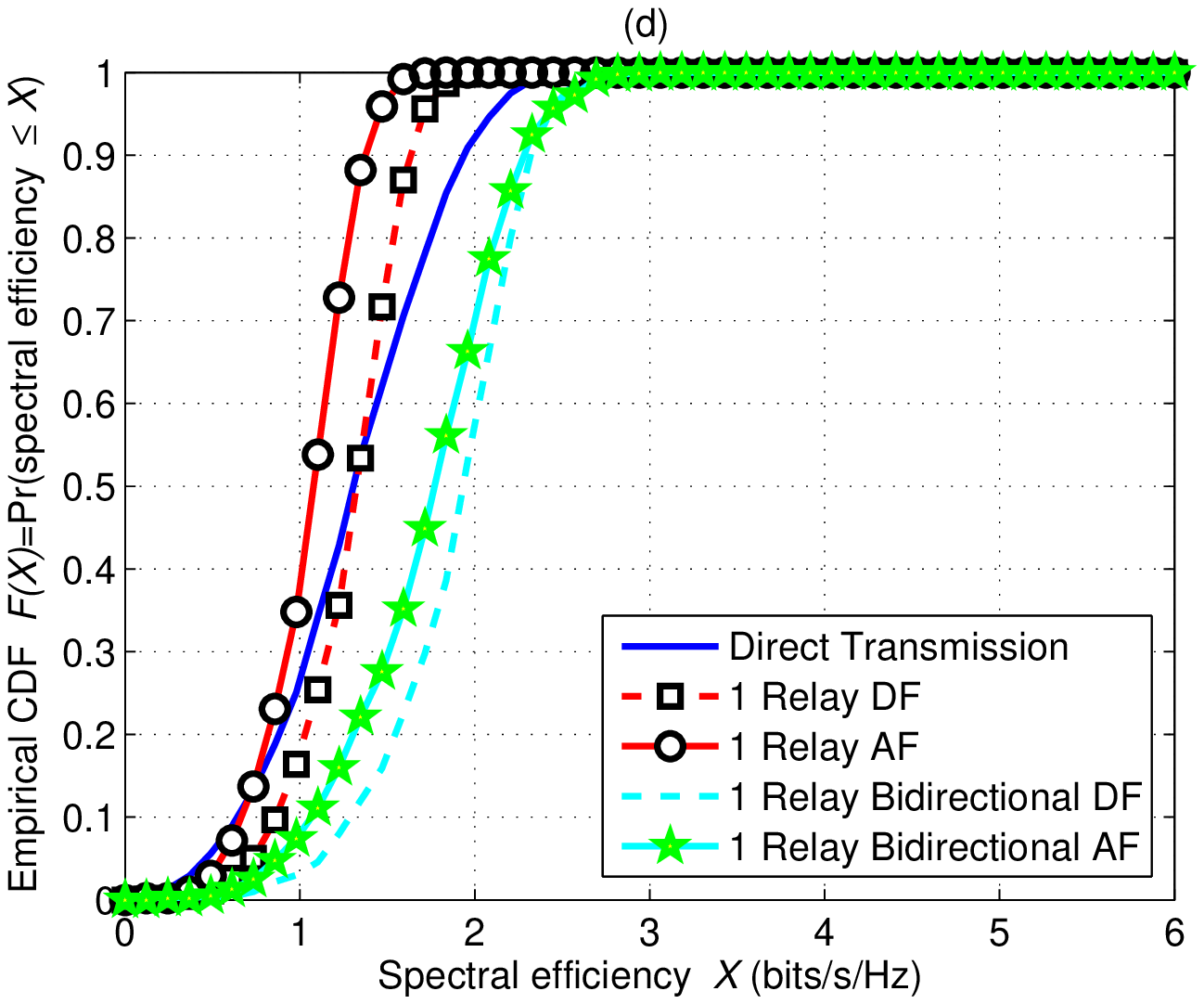}
\end{array}$
\end{center}
\caption{Comparisons of relaying technologies with direct transmission: (a) Spectral efficiency when direct transmission is available, (b) Spectral efficiency when direct transmission is blocked, (c) Empirical spectral efficiency cumulative distribution function when direct transmission is available, and (d) Empirical spectral efficiency cumulative distribution function when direct transmission is blocked.}
\label{fig6}
\end{figure}

It can be seen from Fig.~\ref{fig6}(a) and (b) that, despite the blocked direct transmission, the bidirectional relay strategies are always superior to direct transmission when the transmission distance is relatively long. This phenomenon arises from the facts that the bidirectional relay strategies exchange information using fewer time slots when compared to the unidirectional relay strategies, and that the bidirectional relay strategies do not use the direct link for transmitting information as the two end nodes are working in the half-duplex mode. The bidirectional DF is inferior to direct transmission when the transmission distance $L$ is less than $35$ meters. This is because, when using V-BLAST, we treat the signal from another node as interference (it becomes stronger as the transmission distance becomes shorter), thereby decreasing the information rate of the whole system. Furthermore, we can see from Fig.~\ref{fig6}(c) and (d) that even though the bidirectional relay strategies are not as reliable as the conventional relaying strategies,  they can achieve higher spectral efficiency regardless of the availability of direct transmission link.

\section{Conclusions}
\label{section5}

In this article, we have discussed the challenges of wireless communication technologies when applied to the smart grid.  In order to improve the coverage, the spectral efficiency, and the reliability of smart grid communications, we have investigated several potential relaying strategies, e.g., collaborative beamforming for multiple-relay systems, and bidirectional relaying for information-exchange systems. It has been shown that all relaying strategies can improve the reliability of smart grid communications thanks to the spatial diversity. For DR applications which require two-way
information flow, the spectral efficiency and the coverage of smart grid communications can be improved by using bidirectional relaying strategies. For SCADA applications where the data flow is mostly in one direction, two-relay systems can only achieve marginal improvement over a single relay system at the expense of increased implementation complexity.

\section*{Acknowledgement}

H. Sun and A. Nallanathan acknowledge the support of the UK Engineering and Physical
Sciences Research Council (EPSRC) with Grant No. EP/I000054/1. B. Tan, J. Jiang and J. S. Thompson acknowledge support from
the Scottish Funding Council for the Joint Research Institute in Signal and Image Processing between the University of Edinburgh and
Heriot-Watt University, as part of the Edinburgh Research Partnership in Engineering and Mathematics (ERPem). H. V. Poor acknowledges the support of the U. S. National Science Foundation under Grants CNS-09-05086, CNS-09-05398 and CCF-10-16671.

\bibliographystyle{IEEEtran}

\begin{thebibliography}{}
\providecommand{\url}[1]{#1}
\csname url@rmstyle\endcsname
\providecommand{\newblock}{\relax}
\providecommand{\bibinfo}[2]{#2}
\providecommand\BIBentrySTDinterwordspacing{\spaceskip=0pt\relax}
\providecommand\BIBentryALTinterwordstretchfactor{4}
\providecommand\BIBentryALTinterwordspacing{\spaceskip=\fontdimen2\font plus
\BIBentryALTinterwordstretchfactor\fontdimen3\font minus
  \fontdimen4\font\relax}
\providecommand\BIBforeignlanguage[2]{{%
\expandafter\ifx\csname l@#1\endcsname\relax
\typeout{** WARNING: IEEEtran.bst: No hyphenation pattern has been}%
\typeout{** loaded for the language `#1'. Using the pattern for}%
\typeout{** the default language instead.}%
\else
\language=\csname l@#1\endcsname
\fi
#2}}

\end{thebibliography}


\begin{thebibliography}{1}


\bibitem{white}
Trilliant Incorporated, \emph{Wireless WAN for the Smart Grid: Distribution Networking for Today's (and Tomorrow's) Smart Grid Communications Infrastructure}, Redwood City, CA, Mar. 2010.
%
\bibitem{chan}
P. P. Parikh, M. G. Kanabar, and T. S. Sidhu, ``Opportunities and Challenges of Wireless Communication Technologies for Smart Grid Applications,'' in \emph{Proc. IEEE Power and Energy Society General Meeting}, pp.1-7, Minneapolis, MN, USA, July 2010.

\bibitem{han}
J. Wang and V. C. M. Leung, ``Comparisons of Home Area Network Connection Alternatives for Multifamily Dwelling Units,'' in \emph{Proc. 4th IFIP International Conference on New Technologies, Mobility and Security}, pp.1-5, Paris, France, Feb. 2011.

\bibitem{wmn}
V. C. Gungor, B. Lu, and G. P. Hancke, ``Opportunities and Challenges of Wireless Sensor Networks in Smart Grid,'' \emph{IEEE Transactions on Industrial Electronics}, vol.57, no.10, pp.3557-3564, Oct. 2010.

\bibitem{mesh1}
B. Lichtensteiger, B. Bjelajac, C. M\"{u}ller, and C. Wietfeld, ``RF Mesh Systems for Smart Metering: System Architecture and Performance,'' in \emph{Proc. IEEE International Conference on Smart Grid Communications}, pp.379-384, Gaithersburg, MD, USA, Oct. 2010.

\bibitem{mesh2}
G. Iyer, P. Agrawal, E. Monnerie, and R. S. Cardozo, ``Performance Analysis of Wireless Mesh Routing Protocols for Smart Utility Networks,'' in \emph{Proc. IEEE International Conference on Smart Grid Communications}, pp.114-119, Brussels, Belgium, Oct. 2011.

\bibitem{mesh3}
T. Iwao, K. Yamada, M. Yura, Y. Nakaya, A. A. C\'{a}rdenas, S. Lee, and R. Masuoka, ``Dynamic Data Forwarding in Wireless Mesh Networks,'' in \emph{Proc. IEEE International Conference on Smart Grid Communications}, pp.385-390, Gaithersburg, MD, USA, Oct. 2010.

\bibitem{icc}
B. Tan and J. S. Thompson, ``Relay Transmission Protocols for In-Door Powerline Communications Networks,'' in \emph{Proc. IEEE International Conference on Communications Workshops}, pp.1-5, Kyoto, Japan, June 2011.

\bibitem{af}
H. A. Suraweera, P. J. Smith, A. Nallanathan, and J. S. Thompson, ``Amplify and Forward (AF) Relaying with Optimal and Suboptimal Transmit Antenna Selection,'' \emph{IEEE Transactions on Wireless Communications}, vol.10, no.6, pp.1874-1885, June 2011.

\bibitem{model}
``Propagation Data and Prediction Methods for the Planning of Indoor Radio Communication Systems and the Radio Local Area Networks in the Frequency Range 900 MHz to 100 GHz,'' \emph{ITU-R Recommendations}, Geneva, Switzerland, 2001.

\bibitem{poor}
H. Ochiai, P. Mitran, H. V. Poor, and V. Tarokh, ``Collaborative Beamforming for Distributed Wireless Ad Hoc Sensor Networks,'' \emph{IEEE Transactions on Signal Processing}, vol.53, no.11, pp. 4110-4124, Nov. 2005.

\bibitem{nallan1}
B. Jiang, F. Gao, X. Gao, and A. Nallanathan, ``Channel Estimation and Training Design for Two-Way Relay Networks with Power Allocation,'' \emph{IEEE Transactions on Wireless Communications}, vol. 9, no.6, pp.2022-2032, June 2010.

\bibitem{nallan2}
T.-H. Pham, Y.-C. Liang, A. Nallanathan, and H. K. Garg, ``Optimal Training Sequences for Channel Estimation in Bi-Directional Relay Networks with Multiple Antennas,'' \emph{IEEE Transactions on Communications}, vol.58, no.2, pp.474-479, Feb. 2010.

\bibitem{blast}
C.Z.W.H. Sweatman, J. S. Thompson, B. Mulgrew, and P. M. Grant, ``Comparison of Detection Algorithm including BLAST for Wireless Communication using Multiple Antennas,'' in \emph{Proc. IEEE International Symposium on Personal, Indoor and Mobile Radio Communication}, vol.1, pp. 698-703, London, UK, Sept. 2000.

\end{thebibliography}

\newpage

\newpage

\newpage

\newpage

\newpage

\newpage

\end{document}